\documentclass[letterpaper,10pt]{article}
\usepackage{osajnl2}

\usepackage{xspace}
\usepackage[colorlinks=true,citecolor=blue]{hyperref}
\usepackage{amsmath}
\usepackage{amssymb}

\newcommand{\mic}{$\mu$m\xspace}
\newcommand{\as}{\hbox{$^{\prime\prime}$}\xspace}
\newcommand{\degre}{$^\circ$\xspace}

\begin{document}

\title{On-sky multi-wavelength phasing of segmented telescopes with the Zernike phase contrast sensor}

\author{Arthur Vigan$^{1,2,*}$, Kjetil Dohlen$^{1}$ and Silvio Mazzanti$^{1}$}

\address{$^{1}$Laboratoire d'Astrophysique de Marseille, UMR 6110, CNRS, Universit\'e de Provence, 38 rue Fr\'ed\'eric Joliot-Curie, 13388 Marseille Cedex 13, France}
\address{$^{2}$Astrophysics Group, School of Physics, University Of Exeter, Stocker Road, Exeter EX4 4QL, United Kingdom}
\address{$^{*}$Corresponding author: \href{mailto:arthur@astro.ex.ac.uk}{arthur@astro.ex.ac.uk}}
\address{~\\}

\makebox[1.0\textwidth][c]{Received 15 October 2010; revised 5 Feburary 2011; accepted 26 March 2011}

\begin{abstract}
Future Extremely Large Telescopes will adopt segmented primary mirrors with several hundreds of segments. Cophasing of the segments together is essential to reach high wavefront quality. The phasing sensor must be able to maintain very high phasing accuracy during the observations, while being able to phase segments dephased by several micrometers. The Zernike phase contrast sensor has been demonstrated on-sky at the Very Large Telescope. We present the multi-wavelength scheme that has been implemented to extend the capture range from $\pm\lambda/2$ on the wavefront to many micrometers, demonstrating that it is successful at phasing mirrors with piston errors up to $\pm4.0$~\mic on the wavefront. We discuss the results at different levels and conclude with a phasing strategy for a future Extremely Large Telescope. \\

\noindent This version is for astro-ph. The final publised version is available on the OSA website:

\url{http://www.opticsinfobase.org/abstract.cfm?msid=136717} \\

\end{abstract}


\maketitle

\section{Introduction}
\label{sec:introduction}

Large telescopes have always been an important driver for astrophysical discoveries, leading to the design and construction of several 8 to 10~m telescopes. Practical aspects related to manufacturing and handling limit the size of monolithic mirrors to $\sim$8~m, forcing the adoption of segmented primary mirrors to reach larger diameters, e.g. for the Keck telescopes and the \emph{Gran Telescopio Canarias} (GTC). Both of them are constituted of 36 hexagonal segments of $\sim$0.9~m each, resulting in a total diameter of 10~m. The next generation of Extremely Large Telescopes (ELT), which is currently being designed, will naturally adopt segmented primary mirrors to reach diameters from 30 to 42~m. The two main telescopes of the ELT era will be the Thirty Meter Telescope (TMT) \cite{nelson2008}, 30~m in diameter with 492 segments, and the European ELT (E-ELT) \cite{gilmozzi2008}, 42~m with 984 segments.

ELT primary mirrors are subject to various effects that can modify the relative position of the segments, such as the variation of the gravity vector with respect to the telescope, thermal variations, wind and vibrations. This is why the position of each segment must be actively measured and controlled in piston, tip and tilt to reach optical performances of $\sim$10~nm~RMS (Root Mean Square) on the wavefront. The strategy currently foreseen is a combination of edge sensors to monitor the shape of the primary mirror in real-time and of an Optical Phasing Sensor (OPS) to provide zero reference with nanometer precision for these sensors at regular intervals. Indeed, the accuracy of edge sensors is strongly affected by external conditions such as mechanical alignment, temperature and humidity. An absolute reference is thus needed to recalibrate these sensors and keep low cophasing errors. Current OPS designs are based on Shack-Hartmann wavefront sensors similar in their principle to the one used at Keck telescopes \cite{chanan1998,chanan2000} but several other concepts of sensors have been studied.

Requirements faced by these OPS when cophasing a segmented mirror are of two kinds: (1) the coarse initial phasing of the mirror; and (2) the very fine phasing at high precision. The latter is essential to reach the highest optical performances when constraints on the wavefront are exceptionally high, in particular for applications such as direct detection of exoplanets with extreme adaptive optics and coronagraphy on an ELT \cite{cavarroc2006,martinez2008,yaitskova2008}, for which a precision of a few to 30 nanometers RMS must be obtained to reach contrasts of $10^{-8}$ to $10^{-9}$. The former is equally important for the initial phasing of the primary mirror when segments will be installed for the first time on the telescope. It is also foreseen to be faced on a daily basis in the E-ELT for which 2 of the 984 segments will be replaced every day for re-coating. Mechanical alignment of the segments will provide a cophasing precision of $\sim$20~\mic on the wavefront, leaving the remaining correction down to $\sim$10~nm~RMS for a dedicated OPS.

It is in this context that the Active Phasing Experiment (APE) \cite{gonte2008} was designed to test four concepts of OPS in the laboratory and on the Very Large Telescope (VLT). The four sensors are a Shack-Hartmann sensor \cite{mazzoleni2008}, a pyramid sensor \cite{pinna2006}, a curvature sensor \cite{montoya2006} and a Zernike phase contrast sensor \cite{dohlen2004}. The APE bench includes a small (153~mm in diameter) segmented mirror constituted of 61 segments controllable independently in piston, tip and tilt \cite{dupuy2008}, dimensioned to reproduce future ELTs in terms of segment size and gaps between segments when seen from the OPS. The VLT pupil is re-imaged on the ASM in order to create a fake 8~m segmented pupil which can be analyzed and corrected by the four sensors. Since APE is installed on a Nasmyth platform and does not include a pupil derotator, the VLT pupil, and in particular the spiders, rotates with respect to the segmentation pattern created by the ASM. This can lead to phasing errors when the spiders are in a configuration that isolates different areas of the ASM by covering almost aligned segment borders. This problem is specific to APE and will not take place in a real ELT where the spider will remain fixed with respect to segmentation. The absolute position of each segment in APE is referenced by an internal metrology (IM) system \cite{wilhelm2008} that is used to control the ASM with a precision better than 5~nm~RMS several times per second. By means of the IM, the ASM can be set in any configuration of piston, tip and tilt within the limits of $\pm$14.4~\mic in piston on the wavefront, which is the measurement range of the IM.

One of the sensors in APE is the Zernike phase contrast sensor, also called ZErnike Unit for Segment phasing (ZEUS) developed by \emph{Laboratoire d'Astrophysique de Marseille} (LAM) in collaboration with the European Southern Observatory (ESO) and \emph{Instituto de Astrof\'isica de Canarias} (IAC). ZEUS finds its origins in the Mach-Zehnder (MZ) phasing sensor concept \cite{montoya2004,yaitskova2005}, replacing the delicate interferometer setup by a simple phase mask \cite{dohlen2004}. Similar to the Zernike phase contrast approach \cite{zernike1934}, it has been found equivalent to the MZ in terms of performance characteristics. The phase mask, located in the telescope focus, takes the form of a cylindrical depression machined into a glass substrate. Its diameter is close to that of the seeing disk, and its depth corresponds to a phase shift between $\pi/4$ and $\pi/2$ of the light transmitted through the mask compared to that transmitted through the surrounding substrate. Following the mask, a lens projects an image of the telescope pupil onto a detector array. When the observed star is centered on the mask, pupil aberrations appear as intensity variations on the detector as in the classical phase contrast method, but since the mask is larger than the diffraction spot, low-frequency aberrations, in particular those due to atmospheric turbulence, are filtered out. Piston errors between segments, containing important high-frequency components, show up as anti-symmetric intensity variations along segment edges. Enlarging the mask diameter eliminates more of the atmospheric aberrations, but reduces at the same time the width of the anti-symmetric signal, making it more difficult to measure. A similar trade-off is also found for mask depth: a deep mask, giving a $\pi/2$ phase shift, provides a stronger signal than a shallower mask, but the signal also contains a larger symmetric component, which is found to reduce the accuracy of the signal fitting algorithm, particularly for large piston errors. To allow for different observing conditions, stellar magnitudes, observing wavelength, and initial phasing errors, five different phase masks are available in ZEUS, with depth of 100~nm or 175~nm, and diameter of 1.0\as, 1.5\as and 2.0\as \cite{dohlen2006}. All the results reported here were obtained using the 175~nm-thick masks.

Theoretical treatment of ZEUS and fine cophasing performance within the single-wavelength capture range with ZEUS has been studied in Surdej~\emph{et al.} \cite{surdej2010} using observations performed on the APE bench at the VLT. We present here the results obtained during the same observing runs using a closed-loop multi-wavelength phasing scheme that allows reaching a capture range of several micrometers. We first describe our phasing scheme in Sect.~\ref{sec:multi_wavelenth_phasing_scheme}. Then in Sect.~\ref{sec:observations} we present the two observing runs of February and April 2009 with the different configurations that have been tested. In Sect.~\ref{sec:single_wavelength_results} and Sect.~\ref{sec:multi_wavelength_results} we present the results obtained on-sky, respectively when phasing at a single wavelength in open- and closed-loop with large piston errors, and when using our multi-wavelength scheme to estimate and correct these errors. In Sect.~\ref{sec:phasing_strategy_elt} we extrapolate our results to elaborate a phasing strategy for an actual ELT, and finally in Sect.~\ref{sec:comparison_shack_hartmann} we briefly compare our results with a Shack--Hartmann-type sensor.

\section{Multi-wavelength phasing scheme}
\label{sec:multi_wavelenth_phasing_scheme}

In this section we present a multi-wavelength phasing scheme that allows reaching a capture range of several micrometers in piston. Before generalizing to a complete segmented mirror, we consider the ideal situation of two segments having one border in common that we wish to cophase. The two segments respectively have absolute piston values of $p_1$ and $p_2$, generating a phase difference $\Delta\phi = \frac{2\pi}{\lambda}(p_2-p_1) = \frac{2\pi}{\lambda}\Delta p$ at their common border; $\Delta p$ is called the edge piston. Unless specified otherwise, from now on all values will be given in units of $\lambda$ or in nanometers on the wavefront.

\subsection{Capture range in open-loop}
\label{sec:capture_rang_open_loop}

ZEUS normalized signal is constituted of two parts: an anti-symmetric part, which is equivalent to the signal obtained with a Mach-Zehnder interferometer \cite{yaitskova2005}, and a symmetric part, which is specific to the ZEUS signal. The amplitude of the anti-symmetric part is proportional to the sine of the phase difference between the two segments:

\begin{equation}
  \label{eq:signal_amplitude}
  S_{\mathrm{asym}} = A\sin(\Delta\phi),
\end{equation}

\noindent where A is a calibration coefficient which depends on the phase mask physical properties (thickness and diameter) and the observing conditions (seeing). The phasing algorithm developed for ZEUS allows to retrieve the value of $S_{\mathrm{asym}}$ through signal fitting \cite{surdej2010}, an thus to determine the value of $\Delta p$ with:

\begin{equation}
  \label{eq:deltap_determination}
  \Delta p = \frac{\lambda}{2\pi}\Delta\phi = \frac{\lambda}{2\pi}\arcsin\left(\frac{S_{\mathrm{asym}}}{A}\right).
\end{equation}

Since $S_{\mathrm{asym}}$ is a periodic function of $\Delta\phi$, the range of measurable edge pistons is therefore limited to $\pm\lambda/4$ on the wavefront when operating in open-loop. In the following sections we are going to see how this very narrow capture range can be extended by the use of closed-loop and several wavelengths.

As mentioned by Surdej~\emph{et al.}, the symmetric part of ZEUS signal $S_{\mathrm{sym}}$, which is proportional to $\cos(\Delta\phi)$, could be used to extend the capture range to $\pm\lambda/2$. However, on-sky results have shown that the value retrieved for $S_{\mathrm{sym}}$ through signal fitting is not reliable, and thus cannot be used to remove ambiguity on the edge piston estimation, limiting the effective capture range to $\pm\lambda/4$. We note that the amplitude of the symmetric part strongly depends on mask thickness, so the use of a thinner phase mask gives a cleaner, more easily exploitable signal.

\subsection{Closed-loop phasing}
\label{sec:closed_loop_phasing}

\begin{figure}
  \centerline{\includegraphics[width=0.5\textwidth]{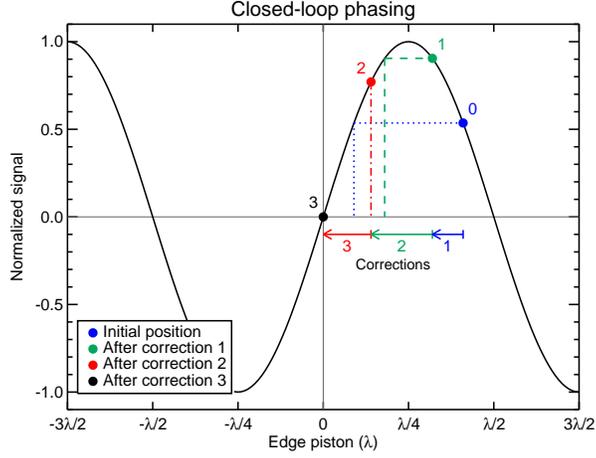}}
  \caption{(Color online) Illustration of closed-loop phasing for an initial edge piston $\Delta p = 0.41\lambda$, i.e. outside of the open-loop capture range. Successive estimations and corrections will always be oriented towards a decrease of the edge piston.}
  \label{fig:zeus_closed_loop_phasing}
\end{figure}

Another possibility to extend the capture range to $\pm\lambda/2$ is to perform closed-loop operation of the OPS. The main idea behind this concept is that for an edge piston $\Delta p \in [\lambda/4\ldots\lambda/2]$, the correction calculated assuming $\Delta p \in [0\ldots\lambda/4]$ will necessarily be oriented towards a decrease of the edge piston. By performing successive measurements and corrections, the edge piston will eventually be brought into the open-loop capture range, and then to zero. This procedure is illustrated in Fig.~\ref{fig:zeus_closed_loop_phasing}, where an edge piston $\Delta p_0 = 0.41\lambda$ is first estimated to $0.09\lambda$, leading to a new edge piston $\Delta p_1 = (0.41-0.09)\lambda = 0.32\lambda$ which is still in $[\lambda/4\ldots\lambda/2]$. The following estimation of $0.018\lambda$ leads to a third edge piston $\Delta p_2 = 0.14\lambda$ which is now within $[0\ldots\lambda/4]$, leading to a final correction that removes the remaining edge piston.

This closed-loop process can be applied for edge steps within $[-\lambda/2\ldots\lambda/2]$, but also around any multiple of $\lambda$, i.e. in $[\lambda(n-1/2)\ldots\lambda(n+1/2)]$, with $n$ an integer. In that case, the closed-loop process will bring the edge piston $\Delta p$ toward $n\lambda$.

\subsection{Dual-wavelength phasing}
\label{sec:dual_wavelength_phasing}

\begin{figure}
  \centerline{\includegraphics[width=0.5\textwidth]{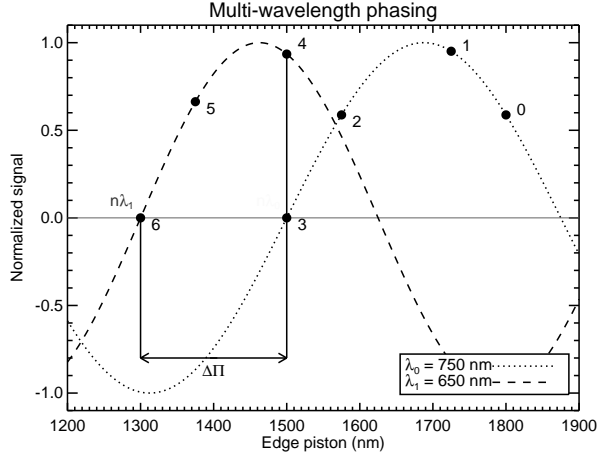}}
  \caption{Multi-wavelength phasing scheme illustrated with $\lambda_0 = 750$~nm and $\lambda_1 = 650$~nm. Phasing is first performed at $\lambda_0$ for the edge piston to converge to $n\lambda_0$, then wavelength is switched to $\lambda_1$ and phasing is performed again until convergence. The $\Delta\Pi$ piston difference between the two phased positions is directly related to the ambiguity $n$ and the two wavelengths (see text for details).}
  \label{fig:zeus_two_wavelengths_phasing}
\end{figure}

If the edge piston $\Delta p$ is larger than $\lambda/2$, the single-wavelength capture range is not sufficient for phasing the two segments with a zero phase error, but only to the closest integer multiple of $\lambda$, leaving an ambiguity of $n\lambda$. The only way to increase the capture range is to use multiple wavelengths to measure and remove this ambiguity. Two-wavelength interferometry, and in particular how to choose the right wavelengths, has already been extensively described in literature \cite{polhemus1973} with several improvements \cite{lofdahl2001,houairi2009}. However, these methods require accurate measurements at two distinct wavelengths. As will be showed in Sect.~\ref{sec:open_loop_performance}, ZEUS measurement accuracy in open-loop on-sky is rather poor, requiring the use of a different method. The multi-wavelength scheme that was adopted for ZEUS \cite{dohlen2006} uses closed-loop phasing at two separate wavelengths $\lambda_0$ and $\lambda_1$ (Simone Esposito, private communication), in order to determine and remove the ambiguity on the edge piston.

Phasing is first performed at $\lambda_0$ until convergence is reached. The edge piston is then $\Delta p_0 = n\lambda_0$, with $n$ an integer. Wavelength is switched to $\lambda_1$, so that a new edge piston $\Delta p_1 = n\lambda_1$ appears:

\begin{equation}
  \label{eq:two_wavelengths}
  \Delta\Pi = \Delta p_0 - \Delta p_1 = n\lambda_0 - n\lambda_1 = n\Delta\lambda,
\end{equation}

\noindent where $\Delta\lambda = \lambda_0-\lambda_1$. While $\Delta p_0$ and $\Delta p_1$ are unknown, we assume that the difference between them, $\Delta\Pi$, is known. The ambiguity $n$ can then be easily determined from Eq.~(\ref{eq:two_wavelengths}):

\begin{equation}
  \label{eq:two_wavelengths_ambiguity}
  n = \frac{\Delta\Pi}{\Delta\lambda}
\end{equation}

In order for the ambiguity $n$ to remain identical for the two wavelengths, Eq.~(\ref{eq:two_wavelengths_ambiguity}) is valid only as long as 

\begin{equation}
  \label{eq:two_wavelengths_validity}
  |\Delta\Pi| < \frac{\bar{\lambda}}{2},
\end{equation}

\noindent with $\bar{\lambda} = (\lambda_0+\lambda_1)/2$. With the approximation that $\Delta p \simeq n\bar{\lambda}$, Eq.~(\ref{eq:two_wavelengths_ambiguity}) and (\ref{eq:two_wavelengths_validity}) lead to the following condition of validity:

\begin{equation}
  2|\Delta p| \lesssim \frac{\bar{\lambda}^2}{\Delta\lambda}
\end{equation}

 For simplicity, we note $\Lambda = \bar{\lambda}^2/\Delta\lambda$ the synthetic wavelength, which is a function of only the two wavelengths $\lambda_0$ and $\lambda_1$. The new multi-wavelength capture range in closed-loop is then equal to $\pm\Lambda/2$.

The multi-wavelength phasing is illustrated in Fig.~\ref{fig:zeus_two_wavelengths_phasing}. In practice, $\Delta\Pi$ is determined by performing phasing of the segments at $\lambda_1$ while recording the successive corrections that are applied on individual segments until convergence is reached at $\lambda_1$. Ambiguity is then determined with Eq.~(\ref{eq:two_wavelengths_ambiguity}) for every segment and finally removed.

Contrary to other methods \cite{lofdahl2001,houairi2009} which require to carefully choose the wavelengths, the choice here is mainly driven by the difference $\Delta\lambda$, over which the capture range depends. Two close wavelengths will provide a wide capture range but will be less precise for small piston errors because the piston difference between the two phased positions will be small, and thus more sensitive to errors. On the contrary, more widely separated wavelengths will provide a shorter capture range, but a greater precision for small piston errors.

\subsection{Generalization and practical implementation}
\label{sec:generalization_practical_implementation}

The closed-loop and multi-wavelength schemes from Sect.~\ref{sec:closed_loop_phasing} and \ref{sec:dual_wavelength_phasing} have been illustrated in the case of two segments having one border in common. However, these schemes can be generalized to a complete segmented mirror with hexagonal geometry. As a matter of fact, the edge piston values at the borders of each segment are related to the segment piston of each segment by a system of linear equations. The segment piston defined with respect to a reference segment which has a piston $p = 0$ by definition. The determination of the edge piston at all segment borders then allows deducing the piston of each segment by the resolution of a set of linear equations using Singular Value Decomposition (SVD).

Another important aspect, which has been overlooked in the previous sections, is the measurement noise. In presence of noise, the OPS will never see a completely phased mirror, i.e. the peak-to-valley (PtV) and RMS values of the measured piston errors will never be zero. It is then necessary to define a convergence criterion, which tells if the mirror is phased from the point of view of the OPS. This criterion has been defined as thresholds $T_{\mathrm{PtV}}$ and $T_{\mathrm{RMS}}$ on the PtV and RMS values of the measured piston errors: for the mirror to be considered as phased, the PtV and RMS must both be lower than their respective threshold. Their exact value is not critical, they must simply be tight enough to phase the mirror with sufficient accuracy, but loose enough to prevent oscillating around the phased configuration because of measurement noise. The optimal values depend on several parameters such as observing conditions, wavelength and phase mask properties. In practice, we used previous results obtained in similar conditions to choose appropriate values, which were in general around $\sim$20 nm for $T_{\mathrm{RMS}}$ and $\sim$50 nm for $T_{\mathrm{PtV}}$. Further study and observations would be required to define an automatic adjustment procedure for these values.

The calibrations for the multi-wavelength scheme leading to the normalized image are identical to that of the single-wavelength scheme \cite{surdej2010}: a dark frame and a reference image taken without the phase mask. For the latter, it is sufficient to off center the mask by a few arcseconds to eliminate its influence on the signal at the segment edges. It is necessary to acquire such an image in each filter used for the multi-wavelength scheme.

\section{Observations}
\label{sec:observations}

ZEUS was tested on-sky during a total of 6 nights from December 2008 to April 2009. A small part of two half-nights ($\sim$6 hours in total) was dedicated to study multi-wavelength phasing of large piston errors, respectively in February and April 2009. Different random configurations were applied on the ASM ranging from 600 to 8000~nm~PtV on the wavefront, outside of the single-wavelength capture range of ZEUS. All these observations were performed on bright stars with magnitudes comprised between V~=~4.0 and V~=~5.0, to avoid being in a noise-limited phasing regime. Investigations of performance in photon starving regime have been presented by Surdej~\emph{et al.}

In February 2009, the 600, 1200 and 2000~nm~PtV random configurations were tested with narrow-band filters at $\lambda_0 = 750$~nm ($\Delta\lambda = 40$~nm) and $\lambda_1 = 650$~nm ($\Delta\lambda = 40$~nm), producing a multi-wavelength capture range of $\pm$4.9~\mic. Observing conditions were average, with seeing varying between 0.7\as and 1.2\as. The diameter of the phase mask used for the observations was 1.5\as and its thickness was 175~nm. 

In April 2009, the 2000, 6000, 7000 and 8000~nm~PtV random configurations were tested at $\lambda_0 = 800$~nm ($\Delta\lambda = 40$~nm) and $\lambda_1 = 750$~nm, producing a capture range of $\pm$12.0~\mic. Observing conditions were good, with a seeing disk around 0.7\as, leading to the use of a smaller mask of 1.0\as with a thickness of 175~nm. 

\section{Single wavelength results}
\label{sec:single_wavelength_results}

\begin{figure*}
  \centerline{\includegraphics[width=1.0\textwidth]{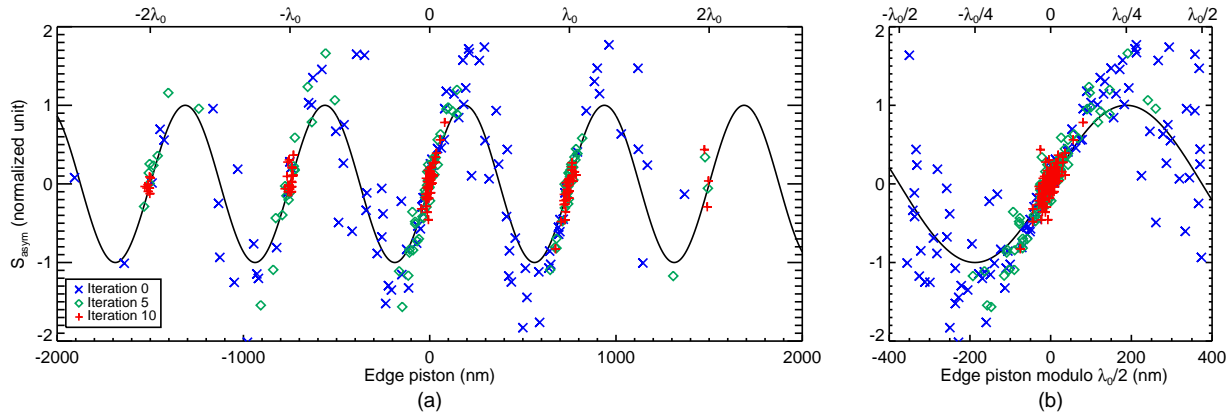}}
  \caption{(Color online) ZEUS anti-symmetric part of the signal, $S_{\mathrm{asym}}$, as a function of edge piston on the wavefront given by the IM (a), and the same measurements folded within $\pm\lambda_0/2$ (b), for the 2000~nm~PtV random configuration on the ASM at $\lambda_0 = 750$~nm. Measurements are given for three iterations during closed-loop phasing. Iteration 0 corresponds to the initial unphased configuration. The ASM is partially phased at iteration 5, and completely phased at iteration 10 according to ZEUS. Borders covered by the VLT pupil and spiders have been removed (see text for details).}
  \label{fig:zeus_borders_2000nm}
\end{figure*}

\begin{figure}
  \centerline{\includegraphics[width=0.5\textwidth]{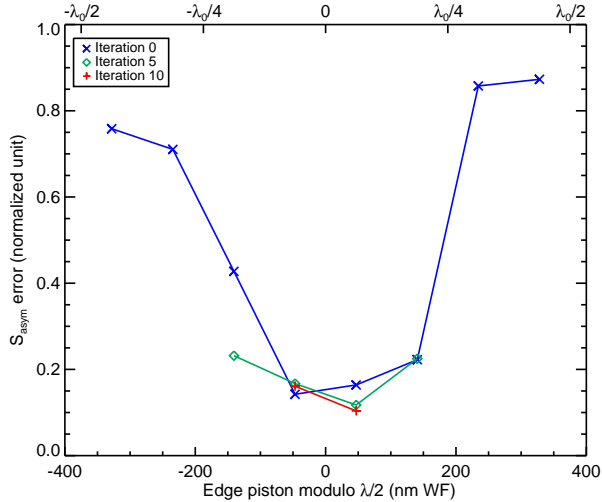}}
  \caption{(Color online) Standard deviation of the error on the determination of $S_{\mathrm{asym}}$ in bins of $\lambda_0/8$ for the same ASM configuration and iterations as Fig.~\ref{fig:zeus_borders_2000nm}. Only bins with more than 5 available points have been represented, but most bins have at least 10 points, thus providing reliable statistics.}
  \label{fig:zeus_borders_stat_2000nm}
\end{figure}

In this section we present the results obtained both in open-loop and closed-loop at a single wavelength. We focus on the analysis of the results at the level of segment borders, where the normalized signal is measured and converted into edge piston.

\subsection{Open-loop performance}
\label{sec:open_loop_performance}

ZEUS shows poor open-loop performance mainly due to large fitting errors in the determination of $S_{\mathrm{asym}}$. Although ZEUS capture range is intrinsically limited to $\pm\lambda/4$ when fitting only the anti-symmetric part of the signal, it is interesting to evaluate the precision with which the signal is fitted in open-loop since it will impact on the convergence speed in closed-loop. Figure~\ref{fig:zeus_borders_2000nm} shows a typical example for a 2000~nm~PtV configuration of the ASM at $\lambda_0 = 750$~nm. Iteration 0 represents the open-loop measurement on this configuration. The data points clearly follow a sinusoidal calibration curve, as expected from Eq.~\ref{eq:signal_amplitude}, but for edge pistons outside of $\pm\lambda/6$, the error can be as high as 50\%. Within this range, the anti-symmetric part $S_{\mathrm{asym}}$ dominates, providing good accuracy on the signal fitting, while outside this range, the symmetric part $S_{\mathrm{sym}}$ becomes more important, and the signal fitting is less accurate.

The loss of open-loop accuracy for edge pistons outside of the quasi-linear part of the sine calibration curve is illustrated on Fig.~\ref{fig:zeus_borders_stat_2000nm}, which shows the standard deviation of the error on the determination of $S_{\mathrm{asym}}$ in bins of $\lambda_0/8$ for the same ASM configuration and iterations as Fig.~\ref{fig:zeus_borders_2000nm}. At iteration 0, all bins are populated with 15 to 20 points (ensuring reliable statistics) and we see that the fitting error clearly increases for edge pistons outside of $\pm\lambda/8$, thus providing accurate measurements only close to zero. This poor open-loop performance, even within the single-wavelength capture range, requires the use of closed-loop phasing to bring all edge pistons close to zero.

\subsection{Closed-loop performance}
\label{sec:closed_loop_performance}

Although open-loop performance is poor, all edge pistons finally converge towards integer multiples of $\lambda_0$ in closed loop. Surdej~\emph{et al.} have shown that convergence is reached with good accuracy for edge pistons within single-wavelength capture range, but we now demonstrate that the same result is obtained for edge pistons around larger integer multiples of $\lambda_0$. We see on Fig.~\ref{fig:zeus_borders_2000nm} that all edge pistons are within $\pm\lambda_0/4$ around a multiple of $\lambda_0$ after 5 iterations, and within $\pm\lambda_0/10$ after 10 iterations, when the ASM is considered to be phased for ZEUS. 

In Fig.~\ref{fig:zeus_borders_stat_2000nm} we see that although the error is large far from zero, all edge pistons finally converge in closed-loop after 10 iterations because the small error in the linear part of the calibration curve allows to retain edge pistons in that area once they are close enough to zero phase error. The same behavior is observed for all other configurations tested on-sky: at the last iteration in single-wavelength, most edge pistons are within $\pm\lambda_0/10$ around a multiple of $\lambda_0$ and the error on $S_{\mathrm{asym}}$ is below 0.2. In the next section we will see how this result translates into the determination of the piston error of the segments.

\section{Multi-wavelength results}
\label{sec:multi_wavelength_results}

As we have seen from the previous section, edge pistons converge in closed-loop towards integer multiple of the wavelength. In this section we are going to see how it leads to the phasing of the ASM at independent wavelengths and how it allows the determination of the piston ambiguity for each segment. From now on we will analyze the results at the level of the segments, i.e. after solving the linear system of equations with SVD that converts the edge pistons into estimated segment piston errors with respect to an arbitrary zero position. These estimates are fed back to the ASM, leading, eventually, to convergence to a phased state, at least in a single-wavelength sense. The absolute piston of the segments is measured by the internal metrology, allowing us to follow the phasing process iteration by iteration and to see the exact position of each individual segment.

\subsection{Convergence at two wavelengths}
\label{sec:convergence_two_wavelengths}

\begin{figure}
  \centerline{\includegraphics[width=0.5\textwidth]{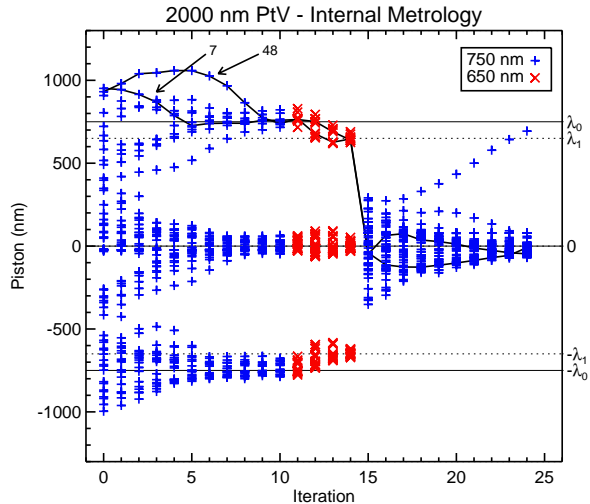}}
  \caption{(Color online) Convergence of all segments using our multi-wavelength scheme for the random 2000~nm~PtV configuration. At the end of iteration 14, the $n\lambda$ ambiguity for each segment is determined and removed, in theory bringing all segments within the single-wavelength capture range. Segments 7 and 48 have been highlighted because they start with similar piston errors, but follow very different paths to reach phasing at $\lambda_0$.}
  \label{fig:zeus_segments_2000nm_all}
\end{figure}

\begin{table}
  \begin{center}
  \caption{Phasing information at $\lambda_0$ and $\lambda_1$}
  \label{tab:phasing_information}
  \begin{tabular}{c|ccc|ccc}
  \hline
  & \multicolumn{3}{c|}{$\lambda_0$} & \multicolumn{3}{c}{$\lambda_1$} \\
  \hline
  PtV  & N$^{\mathrm{a}}$ & RMS$^{\mathrm{b}}$ & Fail$^{\mathrm{c}}$ & N$^{\mathrm{a}}$ & RMS$^{\mathrm{b}}$ & Fail$^{\mathrm{c}}$ \\
  (nm) &       & (nm)    &            &      & (nm)    &            \\
  \hline
   600 & 20    & 72      & 3          & 3     & 59   & 3            \\
  1200 & 24    & 21      & 0          & 4     & 22   & 0            \\
  2000 & 10    & 33      & 0          & 4     & 18   & 0            \\
  2000 & 15    & 10      & 0          & 2     & 17   & 0            \\
  6000 & 14    & 18      & 0          & 6     & 20   & 0            \\
  7000 & 30    & 14      & 0          & 5     & 15   & 0            \\
  8000 & 10    & 75      & 5          & 11    & 13   & 0            \\
  \hline   
  \end{tabular}
  \end{center}
\footnotesize 
$^{\mathrm{a}}$ Number of iterations necessary to reach phasing. \\
$^{\mathrm{b}}$ RMS wavefront error at last iteration for piston errors folded within $\pm\lambda_i/2$. \\
$^{\mathrm{c}}$ Number of segments outside of $n\lambda_i\pm\lambda_i/6$ for which phasing is considered to have failed. \\
\normalsize
\end{table}

The convergence of all edge pistons towards integer multiples of $\lambda$ can equivalently be seen as the convergence of the piston error of each segment towards an integer multiple of $\lambda$. This process is illustrated in Fig.~\ref{fig:zeus_segments_2000nm_all}, which shows the piston error of each segments of the ASM for the 24 iterations that have been performed for the configuration previously showed. This allows to follow the evolution of individual segments or groups of segments. For all ASM configurations that were tested on-sky, the following progression is observed:

\begin{enumerate}
  \item Convergence at $\lambda_0$ from a totally random configuration is somewhat chaotic; segments with similar piston errors at the first iteration can follow very different paths (e.g. segments 7 and 48 in Fig.~\ref{fig:zeus_segments_2000nm_all}).
  \item After a certain number of iterations (the exact number depends on observing conditions and thresholds $T_{\mathrm{PtV}}$ and $T_{\mathrm{RMS}}$) convergence is reached at $\lambda_0$, and wavelength is switched to $\lambda_1$.
  \item The loop is closed at $\lambda_1$ until convergence is reached again. This process takes much less iterations since the mirror is already in an ordered state. Usually all segments converging at $\lambda_0$ will similarly converge at $\lambda_1$.
\end{enumerate}

Convergence was reached for most segments in all ASM configurations tested on-sky, at the exception of a few cases that will be detailed in Sect.~\ref{sec:convergence_problems}. Table~\ref{tab:phasing_information} summarizes all the important phasing information in closed-loop at $\lambda_0$ and $\lambda_1$. In particular, we see that 10 to 30 iterations are necessary to reach phasing at $\lambda_0$, while only 4 to 11 iterations are necessary at $\lambda_1$. In most cases, all segments converge at both wavelengths within $\lambda/6$ or better towards an integer of $\lambda$, translating into 10 to 30~nm~RMS phasing precision on the whole mirror. The final phasing performance is strongly related to the observing conditions, and in particular to the stability of the seeing, but these numbers are comparable to the results obtained in closed-loop around zero \cite{surdej2010}. A larger sample of measurements would be necessary to draw final conclusions on the influence of seeing variations.

\subsection{Ambiguity determination}
\label{sec:ambiguity_determination}

\begin{figure}
  \centerline{\includegraphics[width=0.5\textwidth]{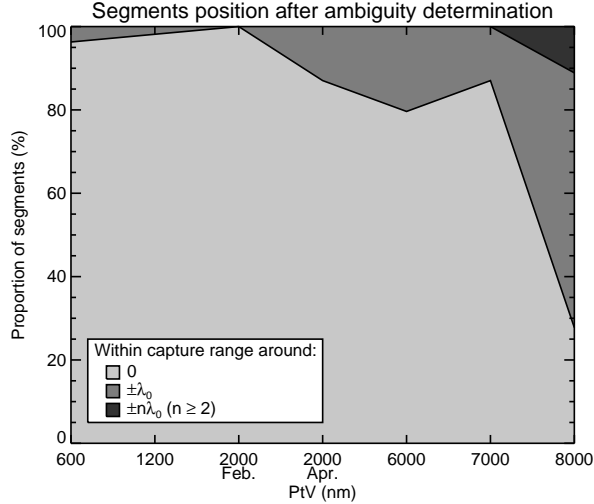}}
  \caption{Proportion of segments brought within the single-wavelength capture range around 0 (light gray), $\pm\lambda_0$ (intermediate gray) or larger multiples of $\pm\lambda_0$ (dark gray) after determination and subtraction of the $n\lambda$ ambiguity using our multi-wavelength scheme for the different ASM random configurations tested on-sky.}
  \label{fig:zeus_convergence_quality}
\end{figure}

We now demonstrate that the accurate closed-loop convergence at both $\lambda_0$ and $\lambda_1$ allows to determine and remove the $n\lambda$ ambiguity on the piston error of the segments to bring them within the single-wavelength capture range. The ambiguity is determined using Eq.~\ref{eq:two_wavelengths_ambiguity} and then removed for each segment. The determination is considered successful for a segment if its piston error is within the single-wavelength capture range around zero after the ambiguity has been removed. The results obtained on-sky are presented in Fig.~\ref{fig:zeus_convergence_quality}, which shows the proportion of segments within single-wavelength capture range around zero, $\pm\lambda_0/2$ and larger multiples of $\pm\lambda_0$ (including segments outside of the multi-wavelength capture range).

The essential result is that for 6 of the 7 ASM configurations tested on-sky, 100\% of the segments are in $[-3\lambda/2\ldots 3\lambda/2]$ after ambiguity removal, and at least 80\% of them are in $[-\lambda/2\ldots\lambda/2]$, which is the single-wavelength capture range. For the 8000~nm~PtV configurations, 2 segments are completely lost because ambiguity determination was 
faulty, and 2 segments are within capture-range around multiples of $\lambda_0$ larger than 2. Using the multi-wavelength scheme again can of course allow to capture these last two segments.

The proportion of segments inside the single-wavelength capture range after ambiguity has been removed decreases for configurations with large initial piston errors, at the benefit of segments that end up close to $\pm\lambda_0$. This is expected since the determination of the ambiguity is quite prone to small errors: the value determined from Eq.~\ref{eq:two_wavelengths_ambiguity} is rounded to the closest integer, so small convergence errors at either $\lambda_0$ or $\lambda_1$ can potentially propagate to produce an error of $\pm1$ on $n$. However, only large convergence errors or segments that did not converge at all could produce errors larger than 1.

Finally, we expect from these results that the number of lost segments will increase for configurations with larger PtV. In particular, configurations with large piston errors are more sensitive to convergence problems that can result in the loss of segments. Configurations having segments close to $\pm\Lambda/2$ or close to the coherence length of the filters will certainly end up with a large proportion of segments either with a very large piston error or within capture range of large multiples of $\lambda_0$. The way to overcome this limitation is to use filters with central wavelengths closer together that will offer a much larger capture range. However, this leads to larger uncertainty in the determination of the ambiguity $n$ (Eq.~\ref{eq:two_wavelengths_ambiguity}), hence of the piston error determination. A second couple of filters with wavelengths more widely separated would then be necessary to recover full phasing precision once all segments are within a few $\lambda$ around the zero position.

\subsection{Systematic effects}
\label{sec:systematic_effects}

In these results, a systematic effect has been removed from the data to provide accurate values. On the ASM, the central segment cannot be actively controlled and is maintained in a fixed position to serve as the zero reference for measuring the piston error of all other segments. While this reference is usable by the OPS in laboratory where there is no telescope pupil covering the ASM, it becomes useless on the VLT where the central reference segment remains hidden behind the secondary mirror. 

However, the Internal Metrology (IM) always sees this reference segment and can thus provide a piston error measurement with respect to the central segment. This has no impact on edge pistons measurements at all borders, which gives a relative positioning information between segments. But when converting the edge piston information into piston error measurements through SVD, the lack of absolute reference introduces a variable offset in all segment pistons determined by the OPS. This changing reference can be referred to as a ``floating'' reference. 

This effect is just an artifact introduced by the IM, and it has no impact on the final single-wavelength phasing performance or on the $n\lambda$ ambiguity determination: ZEUS phases all segments together, simply ignoring the central one, which is the reference for the IM. This is why the floating reference has been estimated and removed from the data presented in Fig.~\ref{fig:zeus_segments_2000nm_all} and Table~\ref{tab:phasing_information}. The estimation is performed at the last iteration of each wavelength using the segments phasing close to zero (zero being defined by the IM). The systematic offset of this group of segments has been measured and removed from all previous iterations at the same wavelength. 

\subsection{Convergence problems}
\label{sec:convergence_problems}

\begin{figure}
  \centerline{\includegraphics[width=0.5\textwidth]{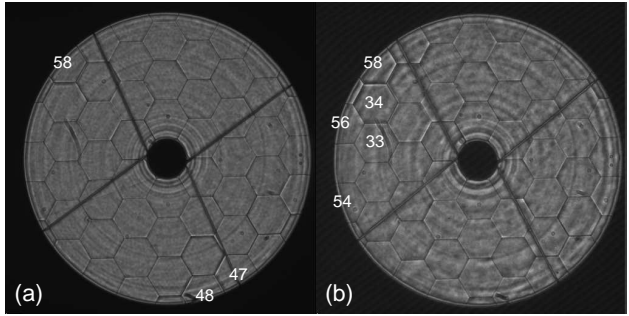}}
  \caption{Segments that failed phasing for the 600~nm~PtV (a) and 8000~nm~PtV (b) random ASM configurations. Images correspond to the last iteration at $\lambda_0$ when the ASM is considered to be phased from the point of view of ZEUS.}
  \label{fig:zeus_bad_segments}
\end{figure}

\begin{figure}
  \centerline{\includegraphics[width=0.5\textwidth]{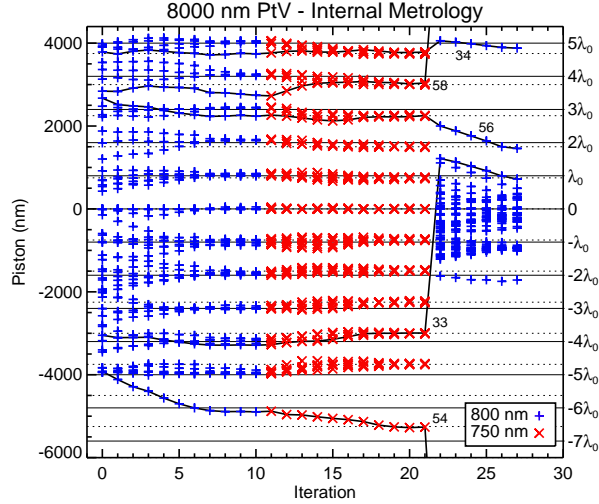}}
  \caption{(Color online) Convergence of all segments using our multi-wavelength scheme for the random 8000~nm~PtV configuration.}
  \label{fig:zeus_segments_8000nm_all}
\end{figure}

Although in most cases all segments converge toward an integer multiple of $\lambda_0$ or $\lambda_1$, there are two notable exceptions visible in Table~\ref{tab:phasing_information}: the 600~nm~PtV configuration at both wavelengths, and the 8000~nm~PtV configuration at $\lambda_0$, for which a few segments failed to converge. These segments have been highlighted in Fig.~\ref{fig:zeus_bad_segments} on the science images obtained for the last iteration at $\lambda_0$. This failure mainly affects segments which are partly covered by the telescope pupil, and where the number of usable borders is generally limited to two or three instead of six, leaving a possible ambiguity on the determination of the piston error of the segment. The spiders can also cover some of the borders and distort the edge piston measurement. Moreover, adjacent segments are influenced by the misbehavior of their neighbor, resulting in groups of segments that fail to converge within the $\pm\lambda/6$ interval.

For the 600~nm~PtV configuration (Fig.~\ref{fig:zeus_bad_segments}(a)), segments 48 and 58 remained stuck close to $-\lambda_0/4$, while segment 47 was influenced by segment 48 and converged only within $\pm\lambda_0/5$ around an integer multiple of $\lambda_0$. For the 8000~nm~PtV configuration (Fig.~\ref{fig:zeus_bad_segments}(b)), a group of four adjacent segments failed to converge at $\lambda_0$: two of them are covered by the pupil and have only 3 borders available for analysis, and the spider is very close to one of these borders. These segments are highlighted in Fig.~\ref{fig:zeus_segments_8000nm_all}, and we can see that segments 34, 56 and 58 remained in between two integer multiples of $\lambda_0$, while segment 33 converged close to $-4\lambda_0$. Segment 54 is different, as it started to diverge at the very first iteration. However, it is interesting to see that all these segments finally converged after wavelength was switched to $\lambda_1$. We can assume that starting from a partially phased ASM and using another wavelength, the uncertainties remaining for these segments were cleared, allowing a final convergence. 

The main problem with these misbehaving segments comes when the $n\lambda$ ambiguity is determined and removed. Since they did not converge properly at one or both of the wavelengths, their value of $n$ can potentially be wrong, sending them outside of the multi-wavelength capture range. This is exactly what happened for segments 54 and 58 in the 8000~nm~PtV configuration: the sign of $\Delta\Pi$, and thus of $n$, was erroneous, and they were sent outside of $\pm\Lambda/2$. For the 600~nm~PtV configuration, since the piston errors are already close to zero, the error on $n$ cannot be larger than 1, which explains why in the end all segments are still within $\pm3\lambda/2$.

\subsection{Summary and limitations}
\label{sec:summary_limitations}

\begin{table*}
  \begin{center}
  \caption{Summary of the multi-wavelength results}
  \label{tab:results_summary}
  \begin{tabular}{cc|cccc|ccccccc}
  \hline
  \multicolumn{2}{c|}{Start} & \multicolumn{4}{c|}{Ambiguity removed} & \multicolumn{7}{c}{End of phasing procedure} \\
  \hline
  PtV & RMS & PtV	& RMS & N$^{\mathrm{a}}$ & Time$^{\mathrm{b}}$ & PtV & RMS & N$^{\mathrm{a}}$ & Time$^{\mathrm{b}}$ & Phased$^{\mathrm{c}}$ & Recov.$^{\mathrm{d}}$ & Lost$^{\mathrm{e}}$ \\
  (nm) & (nm) & (nm) & (nm) & & (min) & (nm) & (nm) & & (min) & & & \\
  \hline
  600	& 163	& 834	& 172	& 25	& 15	& 113	& 25	& 40	& 24	& 61	&     	& 0 \\
  1200	& 339	& 821	& 119	& 29	& 18	& 735	& 94	& 33	& 20	& 60	& 1	& 0 \\
  2000	& 533	& 643	& 153	& 15	& 10	& 763	& 100	& 25	& 15	& 60	& 1	& 0 \\
  2000	& 533	& 1093	& 265	& 18	& 8	& 1867	& 272	& 27	& 11	& 58	& 3	& 0 \\
  6000	& 1668	& 1388	& 318	& 20	& 13	& 895	& 278	& 30	& 16	& 53	& 8	& 0 \\
  7000	& 1698	& 1243	& 265	& 36	& 14	& 1544	& 196	& 45	& 18	& 57	& 4	& 0 \\
  8000	& 2429	& 18802	& 1938	& 22	& 10	& 19676	& 1980	& 28	& 12	& 22	& 37	& 2 \\
  \hline
  \end{tabular}
  \end{center}
\footnotesize 
$^{\mathrm{a}}$ Number of iterations since the beginning of the phasing procedure. \\
$^{\mathrm{b}}$ Total time since the beginning of the phasing procedure. \\
$^{\mathrm{c}}$ Segments in the mono-wavelength capture range ($\pm\lambda/2$). \\
$^{\mathrm{d}}$ Segments within of the multi-wavelength capture range ($\pm\Lambda/2$), i.e. still recoverable. \\
$^{\mathrm{e}}$ Segments outside of the multi-wavelength capture range, i.e. unrecoverable. \\
\normalsize
\end{table*}

In the previous sections we have presented various results for the multi-wavelength scheme of ZEUS. For clarity, different information have been summarized in Table~\ref{tab:results_summary}. They are divided in three categories: the starting configuration, the state after the ambiguity has been determined and removed, and the state when the phasing procedure was ended. The phasing procedure was generally stopped before reaching a new phased state, so the final PtV and RMS values are not necessarily reflecting the ultimate result, but it gives a general view of the number of phased, recoverable and lost segments.

With the use of the multi-wavelength scheme, the piston error of most segments is largely reduced, as can be seen for example in Fig.~\ref{fig:zeus_segments_8000nm_all}. The initial coarse phasing would then be followed by another application of the scheme for a finer phasing with a second pair of wavelengths offering a less extended capture range. Some of the segments have been either lost (only 2 in the configuration with largest piston errors) or did not converge properly. The latter are however recoverable since their piston error still lies within the multi-wavelength capture range.

There are two main limitations inherent to the APE system limiting the performance of ZEUS. The first is that there is no pupil derotator, which means that the spiders are rotating with respect to the segmentation pattern, creating a difference between the images being analyzed and the calibration images taken at the beginning of the phasing procedure. The second is the convergence problems of some segments described in Sect.~\ref{sec:convergence_problems}. It mostly concern segments covered by the telescope pupil edge (or their close neighbors) for which the linear system between the edge pistons and the segment piston error is less over-determined. To better represent a real segmented telescope, the ASM should have been smaller than the VLT pupil.

\section{Phasing strategy for an ELT}
\label{sec:phasing_strategy_elt}

These results can now be analyzed in terms of phasing strategy for a future ELT. We have seen from previous sections that phasing of large piston errors with our multi-wavelength scheme is reliable up to 7000~nm~PtV, and that it could certainly be made more reliable for even larger piston values using filters closer together in wavelength. With a couple of filters at 750~nm and 725~nm, the capture range would be $\pm\Lambda/2 = \pm10.9$~\mic. This capture range would be sufficient to phase an ELT primary mirror which has been mechanically phased with a precision of $\sim$20~\mic, or to recapture a segment which has been changed and reinstalled with equivalent precision.

For an actual ELT, a three-step phasing strategy would appear optimal with a ZEUS-like sensor. First, coarse phasing using a very large capture range in order to remove the largest piston errors should be implemented using filters with wavelengths close together (e.g. 750~nm and 725~nm). Since a correspondingly small bandwidth, typically 10~nm, would be necessary to offer sufficiently long coherence length ($\sim$50~\mic), a bright star (V~$\lesssim$~4) would be required for this step. A second step, using more widely spaced wavelengths (e.g. 750~nm and 650~nm) would bring all segments within the single-wavelength capture range. While correspondingly larger filter bandwidth can be used, a bright star would still be required to limit exposure time. Final phasing is now done using the single-wavelength regime. Here, a wide-band filter can be used, allowing the use of much fainter stars (V~$\lesssim$~10).

Considering the results presented here and by Surdej~\emph{et al.}, phasing of an ELT would require a total of 80 to 100 closed-loop iterations depending on observing conditions in the case of a completely unphased primary mirror. Considering 10 to 20~s exposures, this would require between 15 and 35~min of exposure time. Including target acquisition ($\sim$10~min), calibrations ($\sim$5~min) and computing time ($\sim$20~sec/iteration), this leads to a final estimate of less than 1~h~30 for initial phasing of an ELT. For an already partially phased mirror, only the third step would probably be necessary, consequently reducing the amount of phasing time. Phasing of a single segment with respect to fixed neighbors was not studied, but we believe convergence would be much faster in this case.

Finally, it is important to underline the potential problem of the spiders. In APE, the absence of pupil derotator makes the spider move with respect to the segmentation pattern, sometimes covering several segment borders. The worst configuration occurs when the spiders are almost aligned with borders, thus isolating different parts of the mirror from the point of view of the OPS. The upper right spider of Fig.~\ref{fig:zeus_bad_segments}(a) would be in such a configuration if it was rotated $\sim$5\degre clockwise. Inversely, the best configuration occurs when the spiders are almost orthogonal to segment borders similarly to the lower left spider of Fig.~\ref{fig:zeus_bad_segments}(b). Although it will still prevent good edge piston measurement for these borders, the parts of the mirror lying on each side of the spider will not be isolated from one another because the segments covered have borders on each side of the spider. In order to avoid potential phasing problems, the design of an ELT needs to take into account the size and position of the spiders to avoid as much as possible configurations where different parts of the mirror are isolated.

\section{Comparison with a Shack--Hartmann sensor}
\label{sec:comparison_shack_hartmann}

In this section we briefly compare ZEUS with Shack--Hartmann-type sensors. We base this discussion on the results published by Chanan~\emph{et al.} \cite{chanan1998,chanan2000} on the phasing at the Keck telescopes with their ``broadband algorithm''.

Contrary to a Shack--Hartmann-type sensor, ZEUS does not require any particular alignment between the sensor and the telescope pupil. It is therefore mostly insensitive to alignment variations or optical distortion. Although the success of the Keck phasing sensor proves that such effects are perfectly manageable on a 10~m-class telescope, they can be expected to have a larger impact on the phasing performance for larger telescopes. Also, if continuous phasing is implemented, requiring the use of a guide star whose position in the field varies, pupil distortion may not be constant during the observations. The only constraint for ZEUS is the alignment of the phase mask in the focal plane of the telescope, which needs to be better than 0.1\as on-sky (between 5 and 10\% of the phase mask diameter). Such an alignment could be easily reached with the use of a tip-tilt corrector.

It has been shown that ZEUS phasing accuracy is better than 10~nm~RMS with stars brighter than V~$\simeq$~5 \cite{surdej2010}, which is similar to the accuracy obtained at Keck with the narrow-band phasing algorithm \cite{chanan2000}. When it comes to phasing using a faint star, ZEUS has been shown to reach a precision of 15~nm~RMS for stars of magnitude V~$\simeq$~13. To the best of our knowledge, measurements in similar conditions have not been reported for any other sensor, but this is a criterion for comparison that could potentially be discriminating. Although the ability to phase with faint stars is not necessary for reducing large piston errors, it is certainly important if a continuous phasing strategy is foreseen. For the phasing of large piston errors, it seems that ZEUS is very similar to the Keck Shack--Hartmann sensor in terms of performance and time required for phasing. Chanan~\emph{et al.} report reducing 30~\mic errors in approximately 2~hours. A explained in Section.~\ref{sec:phasing_strategy_elt}, such errors could be reduced with ZEUS in less than 1~h~30 (including calibration) using our multi-wavelength scheme.

\section{Conclusions}
\label{sec: conclusions}

Cophasing of segmented primary mirrors is required to reach high wavefront quality of future ELTs. Given the large number of segments, is it necessary to be able to phase segments with piston errors of several micrometers and to reach a phasing precision of a few nanometers. Moreover, this phasing must be executed simultaneously for all segments. In this work, we have demonstrated on-sky the use of the Zernike phase contrast sensor for phasing of large piston errors using a multi-wavelength scheme. Performing closed-loop phasing at two close wavelengths it is possible to determine and remove the $n\lambda$ ambiguity on the piston of each segments.

Although the open-loop performance is poor due to large errors in the signal fitting procedure outside of $\lambda/4$ around each integer multiples of $\lambda$, the good precision of the fitting close to zero, where the anti-symmetric part of the signal dominates, allows us to reach convergence in closed loop after 10 to 30 iterations at the first wavelength, and in 3 to 11 at the second wavelength. In most cases, all segments phase within $\pm\lambda/6$ around an integer multiple of $\lambda$, so that the $n\lambda$ ambiguity can be estimated accurately. We have demonstrated that for all the ASM configurations tested on-sky, at least 90\% of the segments are within $\pm3\lambda/2$ after ambiguity estimation, and for all configurations except the one with the largest PtV, 80\% of the segments are inside $\pm\lambda/2$, i.e. inside the single-wavelength capture range around zero. Problems directly related to the APE experiment, such as the lack of a derotator, are identified as possible reasons for segments failing to converge properly.

We have also proposed a phasing strategy for an ELT with a ZEUS-like sensor. Using different couples of filters, very large piston errors could eventually be reduced in less than 1~h~30. We have also underlined important problems related to the spider position with respect to the segmentation pattern, and we advocate a configuration where the spiders are perpendicular to segment borders. Such a configuration would avoid isolating different parts of the mirror, resulting in independently phased areas. Finally, we conclude that ZEUS offers similar performances to a Shack--Hartmann-type sensor, but it certainly much more robust in the sense that it does not require any particular alignment with the pupil of the telescope. While this is not problematic in the case of a 10~m telescope, this would certainly be a clear advantage in the 30 to 42~m pupil.

\section*{Acknowledgment}

The APE experiment was part of the ELT Design Study and has been supported by the European Commission, within Framework Programme 6, contract 011863. Mechanical design and manufacturing of the ZEUS sensor were provided by IAC. ESO provided instrument control, most of the signal analysis software, and was for the sensor integration and exploitation within APE. LAM maintained overall responsibility for the development of the ZEUS sensor and provided conceptual design, optical design, phase masks, and alignment and validation in the laboratory, as well as part of the signal analysis software. \\

A. Vigan acknowledges support from a Science and Technology Facilities Council (STFC) grant (ST/H002707/1). We are grateful to Patrick Lanzoni, Marc Ferrari and Maud Langlois at LAM for their various contributions to the ZEUS instrument. We also wish to thank Marcos Reyes and the opto-machanical team at IAC, as well as Isabelle Surdej, Natalia Yaitskova, Fr\'ed\'eric Gonte and Fr\'ed\'eric Derie at ESO. More generally we thank all the people from the APE experiment who have contributed to these results.

%
%


\begin{thebibliography}{10}
\newcommand{\enquote}[1]{``#1''}

\bibitem{nelson2008}
J.~{Nelson} and G.~H. {Sanders}, \enquote{{The status of the Thirty Meter
  Telescope project},} {Proc. SPIE} \textbf{7012} (2008).

\bibitem{gilmozzi2008}
R.~{Gilmozzi} and J.~{Spyromilio}, \enquote{{The 42m European ELT: status},}
  Proc. SPIE \textbf{7012} (2008).

\bibitem{chanan1998}
G.~{Chanan}, M.~{Troy}, F.~{Dekens}, S.~{Michaels}, J.~{Nelson}, T.~{Mast}, and
  D.~{Kirkman}, \enquote{{Phasing the Mirror Segments of the Keck Telescopes:
  The Broadband Phasing Algorithm},} \ao \textbf{37}, 140--155 (1998).

\bibitem{chanan2000}
G.~{Chanan}, C.~{Ohara}, and M.~{Troy}, \enquote{{Phasing the Mirror Segments
  of the Keck Telescopes II: The Narrow-band Phasing Algorithm},} \ao
  \textbf{39}, 4706--4714 (2000).

\bibitem{cavarroc2006}
C.~{Cavarroc}, A.~{Boccaletti}, P.~{Baudoz}, T.~{Fusco}, and D.~{Rouan},
  \enquote{{Fundamental limitations on Earth-like planet detection with
  extremely large telescopes},} A\&A \textbf{447}, 397--403 (2006).

\bibitem{martinez2008}
P.~{Martinez}, A.~{Boccaletti}, M.~{Kasper}, C.~{Cavarroc}, N.~{Yaitskova},
  T.~{Fusco}, and C.~{V{\'e}rinaud}, \enquote{{Comparison of coronagraphs for
  high-contrast imaging in the context of extremely large telescopes},} A\&A
  \textbf{492}, 289--300 (2008).

\bibitem{yaitskova2008}
N.~{Yaitskova}, \enquote{{Adaptive optics correction of segment aberration},}
  Journal of the Optical Society of America A \textbf{26}, 59--+ (2008).

\bibitem{gonte2008}
F.~{Gonte}, C.~{Araujo}, R.~{Bourtembourg}, R.~{Brast}, F.~{Derie},
  P.~{Duhoux}, C.~{Dupuy}, C.~{Frank}, R.~{Karban}, R.~{Mazzoleni},
  L.~{Noethe}, I.~{Surdej}, N.~{Yaitskova}, R.~{Wilhelm}, B.~{Luong},
  E.~{Pinna}, S.~{Chueca}, and A.~{Vigan}, \enquote{{Active Phasing Experiment:
  preliminary results and prospects},} Proc. SPIE \textbf{7012} (2008).

\bibitem{mazzoleni2008}
R.~{Mazzoleni}, F.~{Gont{\'e}}, I.~{Surdej}, C.~{Araujo}, R.~{Brast},
  F.~{Derie}, P.~{Duhoux}, C.~{Dupuy}, C.~{Frank}, R.~{Karban}, L.~{Noethe},
  and N.~{Yaitskova}, \enquote{{Design and performances of the Shack-Hartmann
  sensor within the Active Phasing Experiment},} Proc. SPIE \textbf{7012}
  (2008).

\bibitem{pinna2006}
E.~{Pinna}, S.~{Esposito}, A.~{Puglisi}, F.~{Pieralli}, R.~M. {Myers},
  L.~{Busoni}, A.~{Tozzi}, and P.~{Stefanini}, \enquote{{Phase ambiguity
  solution with the Pyramid Phasing Sensor},} Proc. SPIE \textbf{6267} (2006).

\bibitem{montoya2006}
L.~{Montoya-Mart{\'{\i}}nez}, M.~{Reyes}, A.~{Schumacher}, and
  E.~{Hern{\'a}ndez}, \enquote{{DIPSI: the diffraction image phase sensing
  instrument for APE},} Proc. SPIE \textbf{6267} (2006).

\bibitem{dohlen2004}
K.~{Dohlen}, \enquote{{Phase masks in astronomy: From the Mach-Zehnder
  interferometer to coronagraphs},} EAS Publications Series \textbf{12}, 33--44
  (2004).

\bibitem{dupuy2008}
C.~{Dupuy}, F.~{Gont{\'e}}, and C.~{Frank}, \enquote{{ASM: a scaled down Active
  Segmented Mirror for the Active Phasing Experiment},} Proc. SPIE
  \textbf{7012} (2008).

\bibitem{wilhelm2008}
R.~{Wilhelm}, B.~{Luong}, A.~{Courteville}, S.~{Estival}, and F.~{Gont{\'e}},
  \enquote{{Optical phasing of a segmented mirror with sub-nanometer precision:
  experimental results of the APE Internal Metrology},} Proc. SPIE
  \textbf{7012} (2008).

\bibitem{montoya2004}
L.~{Montoya}, \enquote{{Applications de l'interf\'erom\`etre de Mach-Zehnder au
  cophasage des grands t\'elescopes segment\'es},} Ph.D. thesis, {Universit\'e
  de Provence} (2004).

\bibitem{yaitskova2005}
N.~{Yaitskova}, K.~{Dohlen}, P.~{Dierickx}, and L.~{Montoya},
  \enquote{{Mach-Zehnder interferometer for piston and tip-tilt sensing in
  segmented telescopes: theory and analytical treatment},} Journal of the
  Optical Society of America A \textbf{22}, 1093--1105 (2005).

\bibitem{zernike1934}
F.~{Zernike}, \enquote{{Diffraction theory of the knife-edge test and its
  improved form, the phase-contrast method},} MNRAS \textbf{94}, 377--384
  (1934).

\bibitem{dohlen2006}
K.~{Dohlen}, M.~{Langlois}, P.~{Lanzoni}, S.~{Mazzanti}, A.~{Vigan},
  L.~{Montoya}, E.~{Hernandez}, M.~{Reyes}, I.~{Surdej}, and N.~{Yaitskova},
  \enquote{{ZEUS: a cophasing sensor based on the Zernike phase contrast
  method},} Proc. SPIE \textbf{6267} (2006).

\bibitem{surdej2010}
I.~{Surdej}, N.~{Yaitskova}, and F.~{Gonte}, \enquote{{On-sky performance of
  the Zernike phase contrast sensor for the phasing of segmented telescopes},}
  \ao \textbf{49}, 4052--+ (2010).

\bibitem{polhemus1973}
C.~{Polhemus}, \enquote{{Two-wavelength interferometry},} \ao \textbf{12},
  2071--+ (1973).

\bibitem{lofdahl2001}
M.~G. {Lofdahl} and H.~{Eriksson}, \enquote{{Algorithm for resolving 2pi
  ambiguities in interferometric measurements by use of multiple wavelengths},}
  Optical Engineering \textbf{40}, 984--990 (2001).

\bibitem{houairi2009}
K.~{Houairi} and F.~{Cassaing}, \enquote{{Two-wavelength interferometry:
  extended range and accurate optical path difference analytical estimator},}
  Journal of the Optical Society of America A \textbf{26}, 2503--2511 (2009).

\end{thebibliography}
\end{document}